\documentstyle[12pt]{article}
\def\be{\begin{equation}}
\def\ee{\end{equation}}
\def\ba{\begin{eqnarray}}
\def\ea{\end{eqnarray}}

\title{\bf Decoupling or nondecoupling: is that the $R_b$ question?}
\author{D.\ Comelli$^a$ and Jo{\~a}o P.\ Silva$^b$\\
\\
\small $^a$ Departament de F{\'\i}sica Te{\`o}rica and IFIC\\
\small Universitat Val{\`e}ncia -- CSIC \\
\small E-46100 Burjassot (Val{\`e}ncia), Spain \\
\\
\small $^b$ Centro de F{\'\i}sica Nuclear da Univ.\ de Lisboa \\
\small Av.\ Prof.\ Gama Pinto, 2 \\
\small 1699 Lisboa Codex, Portugal\\
\small and\\
\small Centro de F{\'\i}sica\\
\small Instituto Superior de Engenharia de Lisboa.
}
\begin{document}
\maketitle
\begin{abstract}
The top quark is well known for the nondecoupling effects
it implies in $\rho$ and $R_b$.
The recent experimental $R_b$ data exhibits a disagreement
with the SM prediction at more than the $3 \sigma$ level.
It is tempting to explore whether this might be due to nondecoupling
New Physics effects, opposite to those of the top.
We investigate this issue in the context of models with an extra
family of right or left handed, singlet or doublet quarks.
It is shown that, contrary to what one might naively expect,
the nondecoupling properties of a mirror $t^\prime$ do not
have an impact on $R_b$, due to a conspiracy of the mixing angles,
imposed by the requirement that there be no $b$ - $b^\prime$ mixing.
Our analysis agrees with an analysis performed independently, 
which includes this model as a particular case.
\end{abstract}

\newpage

\section{Introduction}

The high precision experiments on electroweak observables have yielded
spectacular confirmations of the Standard Model,
including in the structure of radiative corrections.
In particular, the experiments performed on the $Z$ resonance at LEP have
probed the couplings of the $Z$ to leptons and quarks.
Here, LEP has found significant deviations from the SM predictions
on the observed ratios $R_b = \Gamma_b/ \Gamma_{had}$ and 
$R_c = \Gamma_c/ \Gamma_{had}$ \cite{reviews,Lan95}.
The former presents a $3.7 \sigma$ deviation from the SM value
($R_b^{SM} = 0.2156$ for a mass of the top $m_t=174$ GeV), if $R_c$
is used as a  free parameter to fit the data.
Conversely, one finds $R_b = 0.2205 \pm 0.0017$,
that is $3 \sigma$ away from the SM prediction,
if $R_c$ is fixed at its SM value.
This discrepancy might be the first window into Physics beyond the SM.
Since $R_b$ exhibits at present a bigger deviation,
we shall concentrate on it.
There are several features that make this decay special:
1) since the bottom is the isospin parter of a `heavy' quark,
and scalar-fermion couplings are typically proportional to fermion masses,
new scalars might give here relevant contributions;
2) the hierarchical structure of the CKM matrix indicates that any
heavier family might couple mostly to the $b$ and $t$ quark;
3) the $Z \rightarrow b \bar{b}$ decay is well known for the nondecoupling
loop contributions that it gets from the top quark \cite{topsquared}.
It turns out that, if one ignored these contributions
(in a blatantly unphysical fashion), the result would be
$R_b(m_t = 0) = 0.220$ \cite{Wel95},
in accordance with experiment!

Recently, there have been many models proposed to solve these discrepancies,
highlighting the first two points.
Some solutions have been sought within well motivated theories like
Extended Technicolour \cite{ETC}, although the solutions are
quite contrived, and Supersymmmetry \cite{SUSY,Sim96}.
In the later, the modifications arise through radiative
corrections\footnote{The simple two Higgs doublet model also
contributes radiatively to $R_b$,
but it only provides a solution when the pseudoscalar mass
is not much larger than 50 GeV, and $\tan{\beta} > 70$ \cite{THDM}.}
, but the relevant parameter space almost disappears in light 
of the recent ``LEP 1.5'' run \cite{Sim96,Ell95}.

There has also been a large number of phenomenological proposals.
In general, we can write the tree level coupling of $Z$ with a
fermion $f$ as
\be
{\cal L}_Z = \frac{g}{c_W} \ Z_\mu
\left[  g_L^f\ \bar{f}_L \gamma^\mu f_L + g_R^f\  \bar{f}_R \gamma^\mu f_R
\right]\ ,
\ee
with
\be
g_\alpha^f = T_3(f_\alpha) - Q(f) s_W^2\ ,
\hspace{7mm}
\alpha = L\, , \, R\ ,
\label{eq:nomixing}
\ee
in the absence of isospin mixing.
This has prompted phenomenological solutions of both problems 
where one allows for tree level mixings of the $b$ and/or $c$
quarks with new quarks of the same charge and different weak
isospin \cite{Bha95,Ma95,Cha96}.
Another possibility arises with the introduction of a new
"hadrophilic" $Z^\prime$ that mixes with the $Z$ \cite{Hol94,Chi96,Alt96}.
For a given choice of parameters, this also allows for an explanation of the 
excess of dijet events observed at CDF.
We note that, in some of the models above one should also worry about the
implications on the oblique radiative corrections.

In this article we stress that the third feature mentioned above
might be the key to the puzzle. Namely, that the disagreement found in
$R_b$ might be signaling us the existence of New Physics through its
nondecoupling effects.
The top quark, recently discovered by CDF \cite{topCDF} and D0 \cite{topD0},
exhibits nondecoupling effect in both $\rho$ and $R_b$.
When the experimental data of the later is confronted with the SM values,
one finds that the radiative corrections (RC) push the SM in the wrong
direction, worsening the problem proportionally to $m_t^2$ \cite{topsquared}.

Here, we will investigate whether this discrepancy might be
explained using the same nondecoupling mechanism that is at work in the SM
but with the opposite sign.
We do so in the context of models with a fourth family of right
or left handed, singlet or doublet quarks.
The mirror quark model stands out as a leading candidate, since
a mirror $t^\prime$ with fixed mixing angle would indeed
produce a nondecoupling effect with the correct sign.
However,
if one requires that there be no tree level $b$ - $b^\prime$ mixing,
a conspiracy of the mixing angles cancels this effect.
In the next section we will briefly discuss nondecoupling effects.
We then turn to the addition of mirror quarks to the SM and 
investigate its consequences.
We finish by presenting our conclusions.

\section{Decoupling and nondecoupling}

As is well known, 
the decoupling theorem states that the physical effects of a 
heavy particle are suppressed at low energies by the inverse powers of the
heavy mass scale, if all the other parameters are held fixed \cite{App75}.
An exception to this theorem occurs naturally in gauge theories with
spontaneous symmetry breaking.
Here, the fermions and scalars often get mass through their
Yukawa couplings with Higgs fields.
When these Higgs fields get (fixed) vacuum expectation values,
we can only increase the mass of those particles increasing the respective
Yukawa coupling.
These large Yukawa couplings may entail a violation of the decoupling
theorem whenever they compensate the heavy mass suppression arising
from the propagator \cite{nondecoupling}.
In the SM the leading RC to the $Z b \bar{b}$ vertex come,
in the Feynman Gauge,
from diagrams involving  top quarks and charged would-be-Goldstone bosons.
These couple very strongly to the fermionic line,
for they are proportional to the Yukawa coupling of the top $y_t$.
However, this fact is not sufficient to get a nondecoupling contribution.

In a very elegant article, Liu and Ng \cite{Liu95} 
have stressed that nondecoupling corrections to the $Z b \bar{b}$ vertex
only occur if the fermion present in the loop transforms chirally under
$SU(2)\times U(1)$, and if its mass is large compared to that of the
exchanged boson.
It has become standard to parameterize both the top and New Physics' impact
on the bottom vertex by \cite{Liu95,Erl95}
\be
g_L^b \rightarrow g_L^b  + \delta g^b_L\ 
\hspace{5mm}, \hspace{5mm}
g_R^b \rightarrow  g_R^b + \delta g^b_R\ .
\ee
These changes are small if one has small tree-level mixing angles or
loop correction.
A recent fit to the electroweak observables yields \cite{Erl95}
\be
\delta g^b_L = - 0.0033 \pm 0.0035
\hspace{5mm} , \hspace{5mm}
\delta g^b_R = + 0.018 \pm 0.013\ ,
\label{eq:limits}
\ee
with a large correlation between these parameters.
To first order, the change in $R_b$ will then be proportional to
$g_L^b \delta g^b_L + g_R^b \delta g^b_R$.
Thereby, a positive $\delta g^b_L$ change will reduce $R_b$, worsening the
problem one already has without such
contributions\footnote{
If $\delta g^b_R=0$ one recovers the case discussed in ref.\ 
\cite{Ber91},
with $\delta_{b-\mbox{vertex}} = 2 \delta g_L^b $,
and in ref.\  \cite{Epsb},
with $\epsilon_b = - 2 {\rm Re} (\delta g_L^b)$.
}.
A similar situation occurs for a negative $\delta g^b_R$ change.

We follow Liu and Ng, and take an interaction of the form 
$ y_F \bar{b}_L \phi^- F_R$ + h.c. , where $\phi$ 
is the would-be-Goldstone boson , $F_R$ an ordinary or new right 
handed quark whose left handed partner, $F_L$, may
transform differently under $SU(2)$.
Obviously, neglecting $m_b$,
one only changes the left handed coupling.
In the $m_F \gg M_Z$ limit, we find
\be
\delta g^b_L 
= 
\frac{\alpha}{4 \pi s_W^2}\ 
\left( \frac{ y_F}{g} \right)^2  
\left[T_3(F_R)-T_3(F_L) \right]\ 
\Delta_0(\frac{m_F^2}{M_Z^2}) \ +\ 
{\rm O}(\frac{M_Z^2}{m_F^2})
\ee
where $\Delta_0(x)= \frac{x}{1-x} +\frac{x}{(1-x)^2} \log x$
(which lies between 0 and $-1$),
confirming the results of ref.~\cite{Liu95}.
In the case of the top quark ($T_3(t_{R})=0$ and $T_3(t_{L})=1/2$)
one finds the well known $m_t^2$ dependence \cite{topsquared},
\be
\delta g_L^{b\ {\rm SM}} \propto
\left( \frac{ y_t}{g} \right)^2  
\left[ - \frac{1}{2} \right] ( -1)
= 
+ \frac{m_t^2}{4 M_W^2}\ .
\label{eq:mt2}
\ee
The fact that this change is positive shows that the nondecoupling
one loop contribution due to the top reduces $R_b$,
worsening the problem considerably,
and leading to the final $3.7 \sigma$ deviation from experiment.

In the case of a new vector-like top' (singlet or doublet),
$T_3(F_{R})=T_3(F_{L})$ and the induced RC are subleading with respect to
the SM one,
\be
\delta g_L^{b\ {\rm vectorlike}} \propto {\rm O}(\frac{M_Z^2}{m_{t'}^2})
\ee
However, a very interesting situation occurs in the presence of
mirror fermions.
In that case,
the isospin quantum numbers of the $t^\prime$
are $T_3(t^\prime_{R})=1/2$ and $T_3(t^\prime_{L})=0$,
generating a contribution of the form
\be
\delta g_L^{b\ {\rm mirror}} \propto -
\frac{m^2_{t^\prime}}{4 M_W^2}\ .
\label{eq:mtprime2}
\ee
which is nondecoupling like the top, but appears naturally with the 
opposite sign.
Notice that, as in the case of the SM, the $\gamma b\bar{b}$ vertex 
is protected from non decoupling contributions by the Ward identities,
as can be checked explicitly performing the substitutions
$g/2 c_w (T_3-2 s_w^2 Q) \rightarrow e Q $ and $ T_3 \rightarrow 0$.
These (far too) simple considerations would lead one to believe
that the addition to the SM of a fourth family of mirror fermions,
in which $t^\prime$ mixes with $t$,
might entail a simple and natural phenomenological solution of the
$R_b$ puzzle.
We shall prove in the next section that this is not so and explain
how that arises as a consequence of the requirement that there be
no tree level $b$ - $b^\prime$ mixing.

\section{The decoupling of mirror fermions}

In this section we present the main features of a model with an extra 
fourth mirror family.
We will concentrate only on the quark sector.
The extra leptons
(present only in order to cancel the anomaly)
are assumed to  decouple completely from the SM ones.
We will follow the notation of Lavoura and Silva 
\cite{Lav93a,Lav93b}
and parameterize the mixing of singlet and doublet quarks
with mixing matrices $V_L$ and $V_R$ in terms of which the
charged boson interactions become
\be
\frac{g}{\sqrt{2}} W_\mu^\dagger \ 
\bar{u}_i \gamma^\mu\ 
{\left[ V_L \gamma_L + V_R \gamma_R  \right]}_{ij} \ d_j
+ {\rm H.c.}\ ,
\ee
where, $i$ ($j$) runs over the number of charge 2/3 (-1/3) quarks,
and $\gamma_{R,L} = (1 \pm \gamma_5)/2$.
Notice that the matrix $V_R$ will exist whenever the right handed
quarks belong to nontrivial multiplets, since the gauge bosons
couple with fermions through the weak isospin. 
In general, these matrices are not unitary but are part of larger
unitary matrices \cite{Lav93a}.
The neutral current interaction in the presence of mixing becomes,
\be
\frac{g}{2 c_w} Z_{\mu} 
\left[ \bar{u}_i \gamma^{\mu}(- \frac{4}{3} s_w^2 + U^L \gamma_L 
+ U^R \gamma_R)_{ij} u_j +
\bar{d}_i \gamma^{\mu}( \frac{2}{3} s_w^2 - D^L \gamma_L 
- D^R \gamma_R)_{ij} d_j \right] \ ,
\ee
where the hermitian mixing matrices $D$ and $U$ represent projection
operators and are given by
($\alpha=L,R$)
\be
U^\alpha = V_\alpha V_\alpha^\dagger,
\hspace{5mm},
\hspace{5mm}
D^\alpha = V_\alpha^\dagger V_\alpha\ .
\ee
It is easy to see that the effective weak isospin of $d_{j L}$ is then given by
the $jj$ component of $- D^L/2$, and similarly for the others.
Thus, in $Z \rightarrow b \bar{b}$, one is probing
\be
\delta g^b_L = (1 - D^L_{bb})/2
\hspace{5mm} , \hspace{5mm}
\delta g^b_R = - D^R_{bb}/2 .
\label{eq:extra}
\ee
Comparing Eqs.~(\ref{eq:limits}) and (\ref{eq:extra}) we see that the
tree level mixing of the $b$ with a $b^\prime$,
which can be either singlet or (lower component of) doublet in
either the left or right hand, will always deepen the problem.
Therefore, one must impose $D^L_{bb}=1$ and $D^R_{bb}=0$.
Of course, by looking at Eq.~(\ref{eq:nomixing}) one can easily
understand this result and see what isospin assignments
must exist in order to circumvent it \cite{Cha96}.

In what follows we will take a fourth family of mirror quarks,
assuming that the hierarchical structure of the CKM matrix is preserved,
so that the mixing of the new $t^\prime$ will occur
predominantly\footnote{
One might also tackle the experimental $R_c$ results 
by mixing the $c$ quark with $t$ and $t^\prime$. In such a context, the
hierarchy of the CKM matrix might also have an interpretation.
We shall not do that here.
} with $t$.
Due to the absence of $b$ mixing, the matrices have a very simple form
\cite{Lav93b}
\be
V_L = \left[ \begin{array}{cc}
             c_1 & 0 \\
             s_1 & 0 
            \end{array} \right] \;\;
    V_R=\left[ \begin{array}{cc}
             0 & c_3 \\
             0 & s_3 
            \end{array} \right] \;\;   
    U^L =\left[ \begin{array}{cc}
             c_1^2 & c_1 s_1\\
             c_1 s_1 & s_1^2 
            \end{array} \right] \;\; 
    U^R=\left[ \begin{array}{cc}
             c_3^2 & c_3 s_3 \\
             c_3 s_3 & s_3^2 
            \end{array} \right]\ .
\ee
Here, the angles $\theta_1$ and $\theta_3$ are those mixing the
left and right handed $Q=2/3$ quarks when one goes from the weak basis
into the mass basis, through
\be
\left[ \begin{array}{cc}
		\frac{v}{\sqrt{2}} \Delta & M_q\\
		M_p	& \frac{v}{\sqrt{2}} \Sigma
	\end{array}
\right]
=
\left[ \begin{array}{cc}
		c_1  & s_1\\
		-s_1 & c_1
	\end{array}
\right]
\ 
\left[ \begin{array}{cc}
		m_t  &  0\\
		0 & m_{t^\prime}
	\end{array}
\right]
\ 
\left[ \begin{array}{cc}
		s_3  & c_3\\
		-c_3 & s_3
	\end{array}
\right]\ .
\label{eq:changebasis}
\ee
Therefore,
\ba
M_q & = & c_1 c_3 m_t + s_1 s_3 m_{t^\prime} = 0\ ,
\label{eq:vectormass}\\
\frac{v}{\sqrt{2}} \Delta & = & c_1 s_3 m_t - s_1 c_3 m_{t^\prime}\ ,
\label{eq:partnermass}
\ea
where the vectorlike doublet mass term $M_q$, that is built with the left
handed third family doublet and the right handed fourth family doublet,
must be zero since we require that $b$ and $b^\prime$ do not mix.
This constraint equation will be crucial in deriving the
decoupling properties.
Also, since $\Delta$ is the mass term of the upper component of the
doublet containing $b_L$, the would-be-Goldstone boson couples $b_L$ to
the physical quarks, proportionally to it.

In the $m_b \simeq 0$ limit, the Lagrangian describing the interaction
with the charged would-be-Goldstone bosons becomes
\be
\frac{\sqrt{2} \phi^-}{v}
\left[ \bar{u}_i X_{ib} \gamma_L b \right] + {\rm H.c.}\ ,
\label{eq:goldstone}
\ee
where
\be
X = (1 - U^R) diag(m_t, m_{t^\prime}) V_L = \frac{v}{\sqrt{2}} \Delta 
\left[ \begin{array}{cc}
		s_3  & 0\\
		-c_3 & 0
	\end{array}
\right]\ .
\label{eq:X}
\ee
If there is no $t - t^\prime$ mixing, $s_1 = 0 = c_3$,
$V_L=1$, and $U^R=0$ recovering the SM result.
Notice the proportionality to $\Delta$, as we had anticipated.

The total leading contribution to $\delta g^b_L $,
including also the top, is
\ba
\frac{8 \pi s_w^2 M_w^2}{\alpha}\ 
\delta g^b_L & = & 
X^\ast_{ib} X_{jb}
\left\{
\left[ U^R/2 - U^L/2 \right]_{ij} m_i m_j C_0(q^2,m_W,m_i,m_j)
\right.
\nonumber\\
&  & \hspace{18mm}
\left.
\left[ U^R/2 - Q(t) s_w^2 \right]_{ij}  \rho_3(q^2,m_W,m_i,m_j)
\right\}
\nonumber\\
& + &
|X_{ib}|^2 
\left[ T_3(\phi^-) - Q(\phi^-) s_W^2
\right] \rho_4(q^2,m_W,m_i,m_j)\ ,
\ea
where the repeated indices $i,j$ are summed from $t$ to $t^\prime$,
and the functions which appear are defined in ref.~\cite{Den91}.
In deriving this result we have used the trivial fact that $U_R X = 0$.
This generalizes the results in refs.~\cite{Liu95,Den91},
for this case in which there is isospin mixing.
For $m_{t,t^\prime}> M_w$ the $\rho_{3,4}$ contributions are subleading 
and the nondecoupling $C_0$ term yields
\ba
\delta g^b_L & \propto & 
- \frac{(c_1 s_3 m_t-c_3 s_1 m_{t^\prime})^2}{4 M_w^2} 
\left[ (c_1^2-c_3^2)(c_3^2-s_3^2)
\right.
\nonumber\\
& & 
\hspace{26mm}
\left.
+ 2 s_3 c_3(c_1 s_1-c_3 s_3)
\frac{m_t m_{t^\prime}}{m_{t^\prime}^2-m_t^2}
\log \frac{m_{t^\prime}^2}{m_t^2} \right]
\label{eq:deltagbL}
\ea
As expected, the final expression is proportional to $\Delta^2$.
Fixing $m_t=175$ GeV would leave three parameters, were it not
for the constraint $m_t c_1 c_3 + m_{t^\prime} s_1 s_3=0$,
imposed by Eq.~(\ref{eq:vectormass}).
Thus $\delta g^b_L$ is a function of only two parameters,
for example, the mixing angles $s_1$ and $c_3$.
The SM result, cf.\ Eq.~(\ref{eq:mt2}),
is correctly reproduced in the limit $s_1=0=c_3$.

Let us look at the impact of Eq.~(\ref{eq:vectormass}) more closely.
It clearly implies that $\theta_1$ and $\theta_3$ must lie in adjacent
quadrants. Using only this fact, one can show that the
coefficient within the squared brackets is always larger than $-1$.
In addition, it can only exceed $0$ marginally, achieving
around $+ 0.125$ when $c_1$ is close to one and $c_3$ is close to $0.85$.
This is easily understood.
In fact, taking $m_{t^\prime}$ to infinity kills the logarithm,
leaving the first term.
This will be positive when $c_1^2 > c_3^2 > s_3^2$. 
For {\it fixed} mixing angles, the overall result would then
exhibit nondecoupling with the correct (negative) sign,
solving the $R_b$ puzzle.
Unfortunately,
Eq.~(\ref{eq:vectormass}) does more than fix the relative signs.
It also implies a relation between the parameters, which we may choose
to write as
\be
s_3^2 = \left[ (\frac{m_{t^\prime} s_1}{m_t c_1})^2 + 1 \right]^{-1}\ .
\ee
For fixed $c_1$, this ends up changing the sign of the decoupling,
as we increase $m_{t^\prime}$.
The limit of $c_1 \ge c_3 \rightarrow 1$
must be taken carefully and we obtain the SM result!

Another interesting limit arises for $m_t=m_{t^\prime}$.
In this case Eq.~(\ref{eq:vectormass}) implies $s_3=-c_1$ and
$c_3=s_1$, and we recover again the SM result.
In fact, any such mixing is allowed, as we can see by looking
back at Eq.~(\ref{eq:changebasis}).
For this case, the mass matrix was already diagonal and
proportional to unity in the weak basis.

Expressing everything in terms of $m_{t^\prime}/m_t > 1$ and $c_1$,
one can show that Eq.~(\ref{eq:deltagbL}) may, at best,
reproduce the SM
\footnote{There is a region for
$c_1 \approx 1$ and $m_{t^\prime}/m_t > 10$,
where the new contribution may cancel,
or even exceed the one from the top.
However, this corresponds to values of the Yukawa couplings $\Delta$
and $\Sigma$, way beyond the perturbative regime.
}.
This occurs whenever $c_1=1$, and also for any case with $m_t=m_{t^\prime}$.
With hindsight, this is a simple consequence of the fact that the $b$
vertex picks up those particles that couple primarily to it,
while $t^\prime$, whose decoupling one would wish to use,
couples primarily to $b^\prime$.
It is tempting to conjecture that such a situation
will occur in any model where this trick is attempted.
Prudence advises that one should wait before
any strong claim is made \cite{Bar96}.

Three possibilities to evade this conclusion come immediately to mind.
One may take the $t^\prime$ as the quark produced at the Tevatron and
have a lighter $t$ \cite{Muk91,Hou95}.
This reduces the $m_t^2$ prefactor ameliorating the problem.
Other possibilities arise taking $m_{t^\prime}/m_t < 1$.
Either $t$ is produced at the Tevatron and $t^\prime$ is lighter
than $175$ GeV, or the $t^\prime$ is the one produced at the Tevatron
and $t$ is heavier than $175$ GeV.
These options already have strong experimental constraints
from direct searches, and we shall not discuss them further
except to point out other nondecoupling properties that must be faced.

In fact,
one must also worry about the nondecoupling effects
present in the oblique radiative corrections \cite{oblique,Pes90}. 
We adopt the $S$ and $T$ parameterization of Peskin and Takeuchi,
for which the last reported constraints are within
$1 \sigma$ of the SM, but tending towards negative values \cite{Lan95}.
Any violation of the custodial symmetry through mass splittings 
among particles inside a multiplet will have an impact on $T$.
In turn, the $S$ parameter is sensitive to the chiral breaking. 
Their expressions for extensions of the SM with the addition of an arbitrary 
number of vectorlike or mirror fermions are given in ref.~\cite{Lav93b}.
For the simple case in which $m_t = m_{t^\prime}$, one finds
\ba
T & = & T^{\rm SM}(m_t,m_b) + T^{\rm SM}(m_t,m_{b^\prime})\ ,
\nonumber\\
S & = & S^{\rm SM}(m_t,m_b) + S^{\rm SM}(m_t,m_{b^\prime})\ ,
\ea
where the first term is the SM contribution and the second has the same
functional dependence but with $m_b$ substituted by $m_{b^\prime}$.
This illustrates in a simple way that $m_{b^\prime}$ cannot be
much larger than $m_{t^\prime}$ for $T$ grows with the difference of the
squared masses.
On the other hand, a fourth family of degenerate fermions yields an
additional $2/(3 \pi)$ contribution to $S$ and is allowed at the
95\% level \cite{Lan95}.
In strict model building,
one might introduce particles in higher multiplets \cite{Dug91}
to reduce $S$, but that lies outside the scope of this article.

\section{Conclusions}

Prompted by the $R_b$ puzzle,
we have analyzed the decoupling properties of an extra $t^\prime$ quark
that runs in the  dominant $Z \rightarrow b \bar{b}$ vertex correction.
This is done in the context of a model with extra left or right handed,
singlet or doublet quarks.
It is well known that a sequential family
produces effects that go in the wrong direction. 
Vectorlike effects are subdominant.

We point out that a mirror $t^\prime$ with fixed mixing angle would exhibit
nondecoupling with the correct sign, apparently solving the problem.
However, when one imposes the absence of tree level $b$ - $b^\prime$ mixing
(that would take $R_b$ in the wrong direction),
the mixing angles conspire to destroy those nondecoupling effects.

It remains to be seen whether complete models may be built with such
nondecoupling New Physics effects.
That would be a very elegant solution to the $R_b$ puzzle. 
It would be similar to the situation that occurred when one knew
that the top quark had to exist due to its nondecoupling effects in $\rho$,
prior to its discovery by CDF \cite{topCDF} and D0 \cite{topD0}.
Our study of the decoupling properties of the mirror $t^\prime$
shows the importance of the mixing angles in extracting such conclusions,
and selecting viable models.

\vspace{10mm}

{\bf Final Note:} After this work was completed we received a comprehensive
independent analysis by Bamert, Burgess, Cline, London, and Nardi \cite{Bam96}.
Their analysis includes this model as a particular case and we agree
with their results. Here we have chosen to concentrate on the nondecoupling
properties from the start and, hence, include the crucial constraint
Eq.~(\ref{eq:vectormass}) at the end.
Since this constraint is common to a large class of models,
they have used it directly in the initial Lagrangian.
This corresponds to noting that the $ib$ components of
$U^R diag(m_t, m_{t^\prime}) V_L$ are proportional to $M_q$,
so that, if one uses Eq.~(\ref{eq:vectormass}) directly on the Lagrangian of
Eq.~(\ref{eq:X}) one reproduces their Eq.~(48) \cite{equality}.

\vspace{10mm}

{\bf \large Acknowledgments}

\vspace{1ex}

We would like to thank A.\ Barroso and L.\ Lavoura for useful advice.
J.\ P.\ S.\ is indebted to A. Eir{\'o} for the facilities provided,
and to L.-F.\ Li for constantly stressing the importance of the decoupling
properties.
The work of D.\ C.\ was funded by the Spanish Comisi{\'o}n
Interministerial de Ciencia y Tecnolog{\'\i}a.
We are greatly indebted to the authors of reference \cite{Bam96},
who have graciously pointed out to us a wrong statement we had originally
made about their article.

\vspace{5mm}


\begin{thebibliography}{99}
\bibitem{equality}
We are indebted to the authors of reference \cite{Bam96}, who have kindly
pointed this out to us.
\end{thebibliography}
\end{document}